\documentclass[%twocolumn,
superscriptaddress,aps,preprintnumbers,amsmath,amssymb,prd,nofootinbib,preprint]{revtex4-1}

\usepackage[dvipdfm]{graphicx}
\usepackage{amsfonts}
\usepackage[mathscr]{eucal}
\usepackage{ascmac}
\usepackage{epstopdf}
\usepackage{dcolumn}% Align table columns on decimal point
\usepackage{bm}% bold math
\usepackage{color}

\newcommand{\bea}{\begin{eqnarray}}  
\newcommand{\eea}{\end{eqnarray}}
\newcommand{\beq}{\begin{equation}}  
\newcommand{\eeq}{\end{equation}}

\newcommand{\lmk}{\left(}  
\newcommand{\rmk}{\right)}
\newcommand{\lkk}{\left[}  
\newcommand{\rkk}{\right]}

\newcommand{\del}{\partial}  
\newcommand{\abs}[1]{\vert{#1}\vert}

\newcommand{\la}{\left\langle} 
\newcommand{\ra}{\right\rangle}

\newcommand{\bear}{\begin{array}}
\newcommand{\eear}{\end{array}}

\newcommand{\dd}{\text{d}}
\newcommand{\MP}{M_*}
\newcommand{\mphi}{m_\phi}

\newcommand{\half}{\frac{1}{2}}

\begin{document}

\preprint{IPMU13-0216}
\preprint{ICRR-Report-667-2013-16}

\title{
Decay rates of Gaussian-type I-balls
and Bose-enhancement effects
in~$3+1$~dimensions
}
\author{Masahiro Kawasaki}
\email{kawasaki@icrr.u-tokyo.ac.jp}
\affiliation{Kavli IPMU (WPI), TODIAS, University of Tokyo, Kashiwa, 277-8583, Japan}
\affiliation{ICRR, University of Tokyo, Kashiwa, 277-8582, Japan}

\author{Masaki Yamada}
\email{yamadam@icrr.u-tokyo.ac.jp}
\affiliation{Kavli IPMU (WPI), TODIAS, University of Tokyo, Kashiwa, 277-8583, Japan}
\affiliation{ICRR, University of Tokyo, Kashiwa, 277-8582, Japan}

%\affiliation{Kavli Institute for the Physics and Mathematics of the Universe, 
%     University of Tokyo, Kashiwa, Chiba 277-8582, Japan}
%\affiliation{Institute for Cosmic Ray Research, 
%     University of Tokyo, Kashiwa, Chiba 277-8582, Japan}

\begin{abstract}
I-balls/oscillons are long-lived spatially localized lumps of a scalar field which may be formed after inflation. In the scalar field theory with monomial potential nearly and shallower than quadratic, which is motivated by chaotic inflationary models and supersymmetric theories, the scalar field configuration of I-balls is approximately Gaussian. If the I-ball interacts with another scalar field, the I-ball eventually decays into radiation. Recently, it was pointed out that the decay rate of I-balls increases exponentially by the effects of Bose enhancement under some conditions and 
%%%%%%%%%%%%%%%%%%%%%%%%%%%%%ver.2%%%%%%%%%%%%%%%%%%%%%%%%%%%%%
%is calculated in $1+1$ dimensions. In this paper, we calculate decay rates of the Gaussian-type I-ball in $3+1$ dimensions and 
a non-perturbative method to compute the exponential growth rate has been derived. In this paper, we apply the method to the Gaussian-type I-ball in $3+1$ dimensions assuming spherical symmetry, and calculate the partial decay rates into partial waves, labelled by the angular momentum of daughter particles.
%%%%%%%%%%%%%%%%%%%%%%%%%%%%%ver.2%%%%%%%%%%%%%%%%%%%%%%%%%%%%%
We reveal the conditions that the I-ball decays exponentially, which are found to depend on the mass and angular momentum of daughter particles and also be affected by the quantum uncertainty in the momentum of daughter particles.
\end{abstract}

%\arxivnumber{1311.0985}

\maketitle

\flushbottom

%%%%%%%%%%%%%%%%%%%%%%%%%%%%%%%%%%%%%%%%%%%%%%%%%%%%%%%%%%%%%%%%
\section{\label{sec1}Introduction}
%%%%%%%%%%%%%%%%%%%%%%%%%%%%%%%%%%%%%%%%%%%%%%%%%%%%%%%%%%%%%%%%

In many real scalar field theories,
there exist long-lived quasi-solitons called I-balls or oscillons~\cite{
Dashen:1975hd, Bogolyubsky:1976nx, Gleiser:1993pt, Kolb:1993hw, 
Copeland:1995fq, Greene:1998pb, McDonald:2001iv, I-ball, Broadhead:2005hn, 
Saffin:2006yk, Hindmarsh:2006ur, Fodor:2006zs, Farhi:2007wj, Hindmarsh:2007jb, epsilon, 
Amin:2010jq, Gleiser:2010qt, Amin:2010xe, Amin:2010dc, Gleiser:2011xj, oscillon2011, Kawasaki:2013hka}
(hereafter we call them as I-balls following Ref.~\cite{I-ball}),
which are spatially localized field condensations.
Although I-balls are not associated with conserved charges,
they are extremely long-lived~\cite{
Segur:1987mg, Graham:2006xs, Gleiser:2008ty, 
Fodor:2008du, Fodor:2009kf, Gleiser:2009ys, previous work, Salmi:2012ta}
due to the existence of an adiabatic invariant,
which is approximately conserved if the potential of scalar field is nearly 
quadratic~\cite{I-ball}.
This adiabatic invariant also reveals the condition for the existence of I-balls.
In Ref.~\cite{I-ball}, it was found that I-balls can be formed
if the potential of a scalar field is nearly and shallower than quadratic.

I-balls are formed in a wide range of cosmological scenarios, including hybrid~\cite{
McDonald:2001iv, Broadhead:2005hn, Gleiser:2011xj} and
chaotic inflation~\cite{oscillon2011}.
In this paper, we consider I-balls in the scalar field theory
with potential of $V(\phi) \propto \phi^{2(1-K)}$ and $0 < K \ll 1$,
which is motivated by chaotic inflation models and supersymmetric theories.
The chaotic inflation model with monomial scalar potential is 
particularly important
since it is simple and avoids the fine tuning of the initial condition for inflation.
The resent upper bound on the tensor-to-scalar ratio~\cite{WMAP, Planck} favors relatively flat potentials $K > 0$.
In inflation with such a flat potential,
the inflaton starts to oscillates after inflation and 
the oscillating inflaton field feels instability, 
which leads to the formation of I-balls~\cite{oscillon2011}.
In addition, the monomial potentials with $\abs{K} \ll 1$ are also motivated by 
supersymmetric theories.
In these models, there are many scalar fields whose potential is given by
$\half m^2 \phi^2 (1 - K \log (\phi^2/M_*^2) ) \approx 
\half m^2 M_*^2 (\phi/M_*)^{2(1-K)}$,
where the logarithmic dependence comes from radiative corrections and $\abs{K} \ll 1$~\cite{EnMc}.
Either sign of $K$ can be realized depending on interactions, and here we consider the case of $K > 0$.
A scalar field with such a potential may have a large vacuum expectation value during inflation
and start to oscillate around the low energy vacuum after inflation~\cite{DRT}.
Soon after the oscillation,
the scalar field feels instability and may form lumps of scalar field condensation.
If the scalar field is a complex scalar field and has a conserved charge,
the lumps of condensation are non-topological solitons called Q-balls,
whose stability is guaranteed by the conserved charge~\cite{Coleman}.
On the other hand, if the scalar field is a real scalar field and has no conserved charge,
the lumps of condensation are I-balls.

%%%%%%%%%%%%%%%%%%%%%%%%%%%%%ver.2%%%%%%%%%%%%%%%%%%%%%%%%%%%%%
When the energy density of I-balls dominates the Universe before they decay, including the case
that inflaton forms I-balls,
they have to decay into radiation before the Big Bang Nucleosynthesis epoch.
Although I-balls can decay into radiation through classical process
without any interactions between I-balls and the other fields
~\cite{Segur:1987mg, Graham:2006xs, Gleiser:2008ty, 
Fodor:2008du, Fodor:2009kf, Gleiser:2009ys, previous work, Salmi:2012ta},
the decay rate through this process is exponentially suppressed.
If I-balls have interaction of $\phi^n$ with $n = 5, 6, \dots$,
they can decay quantum mechanically
via  $(n-2) \to 2$ annihilations~\cite{previous work}.
However, this effect is absent for the I-balls with potential of $V(\phi) \propto \phi^{2(1-K)}$ and $0 < K \ll 1$,
which we consider in this paper.
We thus need to introduce some interaction between I-balls and another light field.
%%%%%%%%%%%%%%%%%%%%%%%%%%%%%ver.2%%%%%%%%%%%%%%%%%%%%%%%%%%%%%

%%%%%%%%%%%%%%%%%%%%%%%%%%%%%ver.2%%%%%%%%%%%%%%%%%%%%%%%%%%%%%
%When an I-ball interacts with another light field,
%the I-ball is expected to decay into radiation.
%\footnote{
%Although I-balls can decay into radiation
%without any interactions between I-balls and the other fields
%~\cite{Segur:1987mg, Graham:2006xs, Gleiser:2008ty, 
%Fodor:2008du, Fodor:2009kf, Gleiser:2009ys, previous work, Salmi:2012ta},
%we neglect this effect
%since the decay rate of I-ball through this process is exponentially suppressed classically
%and is only a power law even in the quantum theory~\cite{previous work}.}
%Naively, one might expect that the decay rate can be estimated from the collection of elementary decay processes.
%%%%%%%%%%%%%%%%%%%%%%%%%%%%%ver.2%%%%%%%%%%%%%%%%%%%%%%%%%%%%%
When an I-ball interacts with another light field,
one might naively expect that the decay rate can be estimated from the collection of elementary decay processes.
Recently, however, it was pointed out that
the effects of Bose enhancement has to be taken into account correctly
when the field which interacts with the I-ball is scalar~\cite{previous work}.
The basic idea is essentially the same as the one in the context of preheating
except for the inhomogeneities of I-ball configuration.
It is well known that
a scalar field interacting with a homogeneously oscillating inflaton feels parametric resonance,
and this parametric resonance leads to explosive reheating called preheating~\cite{preheating}.
This is because
the decay rate of inflaton is proportional to the number of daughter particles by Bose stimulation.
On the other hand, in the case of I-ball decay,
daughter fields escape from the I-ball and the effects of Bose enhancement are weakened.
%%%%%%%%%%%%%%%%%%%%%%%%%%%%%ver.2%%%%%%%%%%%%%%%%%%%%%%%%%%%%%
%Therefore, the effects of Bose enhancement are important
%when particle production is faster than a certain escape velocity 
%from the I-ball~\cite{previous work}.
If particle production is slower than a certain escape velocity from the I-ball,
the I-ball linearly decays only through elementary decay processes.
The effects of Bose enhancement become important
when particle production is faster than the escape velocity~\cite{previous work}.
%%%%%%%%%%%%%%%%%%%%%%%%%%%%%ver.2%%%%%%%%%%%%%%%%%%%%%%%%%%%%%

%%%%%%%%%%%%%%%%%%%%%%%%%%%%%ver.2%%%%%%%%%%%%%%%%%%%%%%%%%%%%%
If the decay rate of I-balls is affected by Bose stimulation,
both the reheating temperature and the process of reheating are altered.
These properties affect many cosmological motivated scenarios.
If one considers Affleck-Dine baryogenesis to account for the baryon density of the Universe, for example,
baryon density is basically proportional to the reheating temperature~\cite{DRT, Affleck-Dine}.
Another example is that
if one considers supersymmetric theories,
gravitino is introduced and its abundance
increases with increasing reheating temperature.
Unless gravitino is heavier than $O(100)$ TeV or lighter than $O(1)$ keV,
overproduction of gravitino spoils the success of the Big Bang Nucleosynthesis
and thus one obtains the upper bound of reheating temperature~\cite{KKM, KKMY, MMY}.
How reheating completes is also an interesting topic in many scenarios,
including non-thermal production of dark matter~\cite{Moroi Randall, GKR} and non-thermal leptogenesis~\cite{AHKY}.
We thus need to know when the decay of I-ball is affected by Bose stimulation
and to determine the decay rate of I-ball.
%%%%%%%%%%%%%%%%%%%%%%%%%%%%%ver.2%%%%%%%%%%%%%%%%%%%%%%%%%%%%%

%%%%%%%%%%%%%%%%%%%%%%%%%%%%%ver.2%%%%%%%%%%%%%%%%%%%%%%%%%%%%%
%Reference~\cite{previous work} proposed a method to calculate the decay rate 
%of I-balls including the effects of Bose enhancement
%and found that I-balls actually decay exponentially by these effects under some conditions.
%In Ref.~\cite{previous work}, Hertzberg used a small amplitude oscillon 
%in 1+1 dimensions.
%In this paper, we improve his method
%and calculate I-ball decay rates in $3+1$ dimensions for Gaussian-type I-balls
%which can be formed in the scalar field theory with the monomial potential 
%of $0< K \ll 1$.
In Ref.~\cite{previous work}, Hertzberg has proposed a method to calculate the decay rate 
of I-balls including the effects of Bose enhancement in general dimensions.
He has applied it to a small amplitude oscillon 
in $1+1$ dimensions as an example
and found that I-balls actually decay exponentially by these effects under some conditions.
In this paper, we apply his method to Gaussian-type I-balls
which can be formed in the scalar field theory with the monomial potential 
of $0< K \ll 1$ in $3+1$ dimensions.
Assuming spherical symmetry, we calculate the partial decay rates of Gaussian-type I-ball
into partial waves, labelled by the angular momentum of daughter particles.
%%%%%%%%%%%%%%%%%%%%%%%%%%%%%ver.2%%%%%%%%%%%%%%%%%%%%%%%%%%%%%
We also reveal the conditions that the decay rate of Gaussian-type I-ball 
increases exponentially by the effects of Bose enhancement and examine its
dependence on the angular momentum and mass of the daughter particle.

This paper is organized as follows.
In the next section, we review a method to calculate I-ball configurations
and derive the Gaussian-type I-ball configuration.
%%%%%%%%%%%%%%%%%%%%%%%%%%%%%ver.2%%%%%%%%%%%%%%%%%%%%%%%%%%%%%
%In section~\ref{sec3}, we improve the method proposed in Ref.~\cite{previous work}
%and then calculate the I-ball decay rate
%in 3+1 dimensions for the Gaussian-type I-ball.
In section~\ref{sec3}, we apply the method proposed in Ref.~\cite{previous work}
to the Gaussian-type I-ball in $3+1$ dimensions
assuming spherical symmetry, and calculate the I-ball decay rate.
In section~\ref{discussions},
we summarise and discuss our results.
%%%%%%%%%%%%%%%%%%%%%%%%%%%%%ver.2%%%%%%%%%%%%%%%%%%%%%%%%%%%%%
Section~\ref{conclusion} is devoted to the conclusion.

%%%%%%%%%%%%%%%%%%%%%%%%%%%%%%%%%%%%%%%%%%%%%%%%%%%%%%%%%%%%%%%%
\section{\label{sec2}Gaussian-type I-ball}
%%%%%%%%%%%%%%%%%%%%%%%%%%%%%%%%%%%%%%%%%%%%%%%%%%%%%%%%%%%%%%%%

In this section, we consider a scalar field $\phi$ 
with  canonical kinetic term and the potential as
\beq
 V(\phi) = 
 \left\{
 \bear{ll}
 \half m_\phi^2 \phi^2, \qquad &\text{ for } \phi \ll M_*, \\
 \frac{m_\phi^2 M_*^2}{2 (1-K)} \lmk \frac{\phi^2}{M_*^2} \rmk^{1-K}, \qquad &\text{ for } \phi \gg M_*,
 \eear
 \right. \label{potential}
\eeq
where $\mphi$ is the mass of $\chi$, and $M_*$ is the crossover scale.
This potential is essentially equivalent to the one used in Ref.~\cite{oscillon2011},
where $\phi$ was considered as inflaton and its mass $m_\phi$ was deduced 
by the amplitude of the power spectrum of curvature perturbations.
It was found that I-balls are formed when
$\half (K - K^2/10) M_{\text{P}} / M_* \gtrsim 10$,
where $M_{\text{P}}$ ($= 1/ \sqrt{8 \pi G_N}$) is the Planck scale.
In this paper,
we do not fix the parameters $m_\phi$ and $\MP$,
and consider the case of $0 < K \ll 1$.
This is also motivated by supersymmetric theories as explained in section~\ref{sec1}.
%%%%%%%%%%%%%%%%%%%%%%%%%%%%%ver.2%%%%%%%%%%%%%%%%%%%%%%%%%%%%%
Note that
the quantum mechanical decay of I-ball through self-interactions
is absent in this theory,
because it is due to self-interactions of $\phi^n$ with $n \ge 5$~\cite{previous work}.
%%%%%%%%%%%%%%%%%%%%%%%%%%%%%ver.2%%%%%%%%%%%%%%%%%%%%%%%%%%%%%

In order to calculate I-ball decay rates including the effects of Bose enhancement,
we have to calculate the field configuration of I-ball $\phi(r, t)$.
For this purpose,
there is a method in which the amplitude of I-balls is expanded 
by a small parameter $\epsilon$
(see~\cite{epsilon} for detail).
However, this method is only applicable to a scalar field theory with polynomial potentials
like $V = \mphi^2 \phi^2/2 + g_3 m_\phi \phi^3/3 + g_4 \phi^4/4 + \cdots$.
Another method is based on an adiabatic invariant, $I$, which I-ball is named after,
and is applicable to the scalar field theory with general potentials including Eq.~(\ref{potential})~\cite{I-ball}.
Moreover, the stability of solutions obtained by the latter method is guaranteed 
by the conservation of the adiabatic invariant, $I$.
Below we derive the configuration of I-ball for the theory with the potential of Eq.~(\ref{potential})
by the latter method and find that the result is approximately Gaussian.

Let us consider a localized scalar field condensation.
When the deviation from the quadratic potential is small (e.g. $\abs{K} \ll 1$ in our case),
we can assume that the time dependence of the field condensation is factorized as
\beq
 \phi(r, t) \approx \Phi(r) \cos(\omega_0 t),
 \label{factorize}
\eeq
where $\omega_0 \simeq \sqrt{V''} \simeq m_\phi$.
This localized condensation is expected to exchange its energy 
with fields in the outer region.
If the time scales of the interactions are sufficiently larger than $\omega_0^{-1}$,
it is proved that a certain adiabatic invariant is conserved in the system averaged over the period of $T = 2 \pi /\omega_0$.
This is also applicable to the case of self-interactions.
When $(\dd / \dd t) \sqrt{ V''(\phi(t))} /\omega_0 \ll \omega_0$ 
(e.g. $\abs{K} \ll 1$ in our case),
the above condition is satisfied and the adiabatic invariant is conserved.
Thus, we consider the system averaged over the period of $T$
and seek the localized scalar field configuration which minimizes
the energy with a constant adiabatic invariant.
This situation is quite similar to that in the Q-ball solution which is obtained 
by minimizing the energy of the localized complex scalar field configuration 
with a constant charge~\cite{Coleman}.
Referring to the case of the Q-ball,
we can derive the I-ball field configuration.

Now we compute the I-ball solution for a given adiabatic invariant.
The adiabatic invariant $I$ is written as
\beq
 I = \frac{1}{\omega_0} \int \dd^3 x \overline{\dot{\phi}^2},
 \label{adiabatic inv}
\eeq
where the overline represents the average over the period of the motion as
\beq
 \overline{Z} \equiv \frac{1}{T} \int_t^{t+T} \dd t' Z(t').
\eeq
Here, we have used the different overall factor of the adiabatic invariant compared with the one used in Ref.~\cite{I-ball}
so that we can interpret $I$ as the number of scalar particles inside the I-ball (see Eq.~(\ref{E and I})).
The scalar field configuration which minimizes the time-averaged energy
at a fixed adiabatic invariant $I$ is obtained by minimizing
\bea
 E_{\omega} &\equiv& \overline{E} + \tilde{\lambda}_0 \lmk I - \frac{1}{\omega_0} \int \dd^3 x \overline{\dot{\phi}^2} \rmk, 
 \label{E_w}\\
 E &=& \int \dd^3 x \lkk \half \dot{\phi}^2 + \half \lmk \nabla \phi \rmk^2 + V \lmk \phi \rmk \rkk, \label{I-ball energy}
\eea
where $\tilde{\lambda}_0$ is a Lagrange multiplier.
From Eq.~(\ref{factorize}),
we can calculate the time-averaged scalar field configurations as
\bea
 \overline{\phi^2} &=& \half \Phi^2 ( \bm{r}), \\
 \overline{\dot{\phi}^2} &=& \half \omega_0^2 \Phi^2 ( \bm{r}),
\eea
and also we obtain 
\beq
 \half \tilde{V} \lmk \Phi \rmk \equiv \overline{V(\phi)} \simeq 
 \left\{
 \bear{ll}
 \frac{1}{4} m_\phi^2 \Phi^2, \qquad &\text{ for } \phi \ll M_*, \\[0.8em]
 \frac{m_\phi^2 M_*^2}{4 (1-K)} \lmk \frac{\Phi^2}{M_*^2} \rmk^{1-K}
 \xi(K), \qquad &\text{ for } \phi \gg M_*,
 \eear
 \right.
\eeq
where $\xi(K) = \frac{2\Gamma(3/2-K)}{\sqrt{\pi}\Gamma(2-K)}\simeq 1+0.39K$.
Using these equations, Eq.~(\ref{E_w}) is written as
\beq
 E_{\omega} = \half \int \dd^3 x \lkk 
 \half \lmk \nabla \Phi \rmk^2
 + \tilde{V} \lmk \Phi \rmk - \half \tilde{\omega}_0^2 \Phi^2
 \rkk + \tilde{\lambda}_0 I, \label{energy}
\eeq
where
\beq
 \tilde{\omega}_0^2 \equiv \omega_0 \lmk 2 \tilde{\lambda}_0 - \omega_0 \rmk.
\eeq
Taking a spherically symmetric ansatz $\Phi( \bm{r})=\Phi(r )$,
we can calculate the radial part of the configuration by solving the following equation:
\beq
 \frac{\del^2}{\del r^2} \Phi + \frac{2}{r} \frac{\del}{\del r} \Phi + \tilde{\omega}_0^2 \Phi - \tilde{V}' \lmk \Phi \rmk = 0, 
 \label{I-ball eom}\\
\eeq
with the boundary condition $\del \Phi / \del r (0 )=0$ and $\Phi ( r \to \infty) = 0$.
Fortunately, this equation is the same as the one to find Q-ball solutions,
which was well investigated in many papers~\cite{Coleman, Ku, KuSh, EnMc, KK}
and is reviewed below.

In solving Eq.~(\ref{I-ball eom}),
we use the following Gaussian ansatz:
\beq
 \Phi(r ) \simeq \Phi_0 \exp[ - r^2 / (2 R^2)],
\eeq
where $\Phi_0$ is the amplitude at the center of the I-ball,
and $R$ is the typical size of the I-ball.
Substituting this ansatz into Eq.~(\ref{I-ball eom}), we obtain
\beq
 \frac{r^2}{R^4} - \frac{3}{R^2} + \tilde{\omega}_0^2
 - m_\phi^2 \lmk \frac{\Phi_0^2}{ M_*^2} \rmk^{-K} e^{K r^2 / R^2} \xi(K)
 = 0.
\eeq
Using $\exp[K r^2/ R^2] \simeq 1 + K r^2 / R^2$ for $\abs{K} \ll 1$
and comparing the coefficients of $r^n$ ($n=0, 2$),
we obtain
\bea
 R &\simeq& 
 \frac{1}{K^{1/2} m_\phi},
  \label{I-ball R}\\
 \tilde{\omega}_0^2 &\simeq& 
 m_\phi^2,
\eea
where we neglect higher-order terms in $K$.
Note that there is no solution if $K < 0$, and thus we assume $0< K$ $( \ll 1)$.
When we substitute the ansatz into Eqs.~(\ref{adiabatic inv}) and (\ref{I-ball energy}),
we obtain the adiabatic invariant and the energy of the I-ball as
\bea
 I &\simeq& \frac{\pi^{3/2}}{2} \omega_0 \Phi_0^2 R^3, \label{I-ball I}\\
 M_I &\equiv& E \simeq \frac{\pi^{3/2}}{2} \omega_0^2 \Phi_0^2 R^3 \simeq \omega_0 I, 
 \label{E and I}
\eea
respectively.
From Eqs.~(\ref{I-ball R}) and (\ref{I-ball I}) $\Phi_0$ is written as 
\beq
 \Phi_0 \simeq \lmk \frac{2}{\pi^{3/2}} \rmk^{1/2} 
 K^{3/4} I^{1/2} m_\phi. \label{phi0} 
\eeq

From Eq.~(\ref{E and I}) and $\omega_0 \simeq m_\phi$,
the energy of the I-ball is given by $M_I \simeq \mphi I$,
and thus
the adiabatic invariant $I$ can be interpreted as the number of scalar particles $\phi$ carried by the I-ball.
We can estimate the adiabatic invariant of the typical I-balls which are formed after inflation in the following way.
After inflation, the scalar field begins to oscillate around the low energy vacuum $\phi=0$
and the oscillating field  feels spatial instabilities, 
which lead to the formation of I-balls.
The most amplified mode of the scalar field is estimated as $k \approx 1/R$ 
by analogy to Q-ball.
Thus, we can estimate the number of scalar fields carried by the typical I-ball as
\beq
 I = \beta \frac{4 \pi R^3}{3} \frac{ \omega_0 \phi_{osc}^2}{2},
 \label{I estimation}
\eeq
where $\omega_0 \phi^2_{osc}/2$ is the number density of the scalar field, and
$\phi_{osc}$ is the amplitude of the scalar field at the onset of oscillation.
We include a factor $\beta$ in order to take into account the delay 
of the I-ball formation from beginning of the oscillation~\cite{MuNa}.
Note that $\beta \sim 10^{-(2-4)}$ in the case of Q-ball formation~\cite{Qgrav}.
From Eqs.~(\ref{I estimation}) and (\ref{phi0}),
we can estimate the typical amplitude of the I-balls formed after inflation.

%%%%%%%%%%%%%%%%%%%%%%%%%%%%%%%%%%%%%%%%%%%%%%%%%%%%%%%%%%
\section{\label{sec3}Method to calculate decay rates of I-balls}
%%%%%%%%%%%%%%%%%%%%%%%%%%%%%%%%%%%%%%%%%%%%%%%%%%%%%%%%%%

In this section, we consider the theory including the I-ball $\phi(r, t)$ 
and another scalar field $\chi$ with the following
mass and interaction terms:
\beq
 \mathcal{L} \supset - \half m_\chi^2 \chi^2 +  \half g A \phi \chi^{2}. \label{interaction}
\eeq
We can assume $A = \omega_0^2 / \Phi_0$ without loss of generality.
This interaction leads to parametric resonance before the formation of I-balls
when the coupling $g$ is sufficiently large,
and the energy of the oscillating scalar field flows into the energy of 
the field fluctuations of $\chi$ ($\chi$-particles)~\cite{preheating}.
If the growth rate of the field $\chi$ is greater than the growth rate of I-balls,
the $\phi$ oscillation is damped without formation of I-balls.
In this paper, we simply assume that the interaction of Eq.~(\ref{interaction}) turns on 
after I-balls are formed
and investigate the properties of I-ball decay through this interaction.

In order to investigate the I-ball decay rate,
we calculate the number density of the daughter field $\chi$ 
in the leading semi-classical approximation,
which is widely used in the context of soliton decay and 
preheating~\cite{preheating, evap, KuLoSh, KY}.
%%%%%%%%%%%%%%%%%%%%%%%%%%%%%ver.2%%%%%%%%%%%%%%%%%%%%%%%%%%%%%
%In particular, the calculation of the decay rate of I-balls in 1+1 dimensions was studied in Ref.~\cite{previous work}.
%In this section, we improve the method to calculate the I-ball decay rate
%in 3+1 dimensions with the assumption of spherical symmetry.
In particular, a non-perturbative method to calculate the decay rate of I-balls in general dimensions
was derived in Ref.~\cite{previous work}.
In this section, we apply the method to calculate the decay rate of I-ball
in $3+1$ dimensions assuming spherical symmetry.
%%%%%%%%%%%%%%%%%%%%%%%%%%%%%ver.2%%%%%%%%%%%%%%%%%%%%%%%%%%%%%
In order to compute the particle creation rate numerically,
we consider a system in a box of volume $4 \pi L^3 / 3$
and discretize momentum space as $k \to n \pi/L$, where $n$ is an integer and runs from $1$ to $N$.
We set $N \approx \mphi L / \pi$ since we know that the momentum far from $\mphi/2$ is irrelevant
for enhanced decay modes by analogy to preheating.

We treat the I-ball $\phi({\bf r}, t)$ as a classical background field in the leading semi-classical approximation.
In this case,
the Heisenberg equation of motion
of the quantum scalar field $\chi$ with the interaction of Eq.~(\ref{interaction}) becomes linear
and thus can be solved.
We consider the spherically symmetric system with the I-ball background field $\phi(r, t)$ at the origin of the coordinate.
Due to the rotational invariance of the system, we can expand the field $\chi$ as
\beq
 \chi \lmk \bm{r}, t \rmk
 = \sum_{l,m} \sum_{p=1}^N \sqrt{\frac{2}{L^3}} \frac{1}{j_{l+1} \lmk \alpha_{l,p} \rmk }
 j_l \lmk k_{l,p} r \rmk Y_l^m(\theta,\varphi) 
 \tilde{f}_{l,m,p} (t),
\label{R-expansion}
\eeq
where $\tilde{f}_{l,m,p}$ are expansion coefficients, $Y_l^m(\theta,\varphi)$ 
are the spherical harmonics,
and $j_l(kr)$ are the spherical Bessel functions.
Here, $\alpha_{l, p}$ are the $p$-th roots of $j_l$:
\beq
 j_l \lmk \alpha_{l, p} \rmk \equiv 0,
\eeq
and we define $k_{l,p} \equiv \alpha_{l, p} / L$.
The spherical Bessel functions satisfy
the following orthogonality integral:
\beq
 \int_0^L j_l \lmk k_{l,p} r \rmk j_l \lmk k_{l,q} r \rmk r^2 \dd r
 = \frac{L^3}{2} \lkk j_{l+1} \lmk \alpha_{l, p} \rmk \rkk^2 \delta_{p q}.
\eeq
Since the annihilation and creation operators are mixed with each other as time evolves,
the expansion coefficient is written as
\beq
 \tilde{f}_{l,m,p} (t) = \sum^N_{q=1} \lkk v_{l,p}^{\ q} (t) a_{l,m,q} + (-1)^m v_{l,p}^{\ q *} (t) a_{l,-m,q}^\dagger \rkk.
 \label{f-expansion}
\eeq
We impose the initial condition as
\bea
 v_{l,p}^{\ q} (0) = \frac{1}{\sqrt{2 \omega_{l,p}}} \delta_{p}^q, \label{initial condition1} \\
 \dot{v}_{l,p}^{\ q} (0) = \frac{-i \omega_{l,p}}{\sqrt{2 \omega_{l,p}}} \delta_{p}^q, \label{initial condition2}
\eea
where $\omega_{l,p}^2 \equiv k_{l,p}^2 + m_\chi^2$.
After the field $\chi$ is quantized,
the coefficients $a$ and $a^\dagger$ become operators
which satisfy the following commutation relations:
\beq
 \lkk a_{l,m,p}, a^\dagger_{l', m', p'} \rkk = \delta_{ll'} \delta_{mm'} \delta_{pp'},
\eeq
and are interpreted as the annihilation and creation operators, respectively.
The quantum vacuum state is defined by $a_{l,m,p} \left\vert 0 \right\rangle = 0$,
which means that the field $\chi$ is absent at $t=0$.
The energy of the field $\chi$ can be calculated from
\beq
 \la 0 \left\vert H_\chi \right\vert 0 \ra = \half \sum_{l,m,p}
 \lkk \sum_{q=1}^N \lmk \abs{\dot{v}_{l,p}^{\ q}}^2 + \omega_{l,p}^2 \abs{v_{l,p}^{\ q}}^2 \rmk
 - \omega_{l,p} \rkk,
 \label{chi energy}
\eeq
where we subtract the zero-point energy.

We assume that
the I-ball is formed instantaneously at $t=0$ and the configuration of the I-ball is given by
\beq
 \phi(r,t) = 
  \Phi(r) \cos\omega_0 t,
\eeq
where $\Phi(r) = \Phi_0 \exp \lkk - r^2 / (2 R^2) \rkk$.
Using the expansions of Eqs.~(\ref{R-expansion}) and (\ref{f-expansion}),
the equation of motion is written by
\beq
 \lmk \frac{\dd^2}{\dd t^2} + k_{l,p}^2 \rmk v_{l,p}^{\ q}
 + g \omega_0^2 \cos\lmk \omega_0 t \rmk \sum_{p'=1}^N \Phi_{p}^{p'} v_{l,p'}^{\ q} 
 = 0,
 \label{eq-discre}
\eeq
where $\Phi_{p}^{p'}$ is defined as
\beq
 \Phi_{p}^{p'} = \int_0^L \dd r \frac{2}{L^3} 
 \frac{r^2 j_l \lmk k_{l,p} r \rmk j_l \lmk k_{l,p'} r \rmk}
      {j_{l+1} \lmk \alpha_{l,p} \rmk j_{l+1} \lmk \alpha_{l,p'}  \rmk} 
 \frac{\Phi(r)}{\Phi_0}.
\eeq

We introduce the $2 N \times 2 N$ matrix $\bm{M}$ as
\beq
 \lmk
 \bear{l}
  v_{l,p}^{\ q}(t) \\
  \dot{v}_{l,p}^{\ q}(t) \\
 \eear
 \rmk
 = 
 \bm{M}(t)
 \lmk
 \bear{ll}
  \frac{1}{\sqrt{2 \omega_{l,p}}} \bm{1}_{N \times N} \\
  \frac{-i\omega_{l,p}}{\sqrt{2 \omega_{l,p}}} \bm{1}_{N \times N} \\
 \eear
 \rmk,
\eeq
where $\bm{1}_{N \times N}$ is the identity matrix of size $N$.
From this and Eqs~(\ref{initial condition1}) and (\ref{initial condition2}),
we obtain the initial condition of $\bm{M}$ as
\beq
 \bm{M}(0) = 
 \lmk
 \bear{ll}
  \bm{1}_{N \times N} \quad &0 \\
  0 \quad &\bm{1}_{N \times N} \\
 \eear
 \rmk,
\eeq
and the equation of motion~(\ref{eq-discre}) can be rewritten as
\beq
 \frac{\dd}{\dd t} \bm{M}(t) = 
 \lmk
 \bear{ll}
  0 \quad &\bm{1}_{N \times N} \\
  \bm{Q} \quad &0 \\
 \eear
 \rmk
 \bm{M}(t), \label{eq-discre2}
\eeq
where $\bm{Q}$ is the $N \times N$ matrix defined by
\beq
 \lmk \bm{Q} \rmk_{p}^{p'} (t) \equiv - \omega_{l,p}^2 \delta_{p}^{p'} - g \omega_0^2 \cos\lmk \omega_0 t \rmk \Phi_{p}^{p'}.
\eeq
From the periodicity $\bm{Q} ( t+T ) = \bm{Q} (t)$, we obtain $M(n T) = M(T)^n$.
In order to extract the modes whose amplitude increases exponentially,
we need to find eigenvalues $e^{\mu_i t}$:
\beq
 \bm{a}_i \bm{M} (T) = e^{\mu_i T} \bm{a}_i,
\eeq
where $\bm{a}_i$ are corresponding eigenvectors.
From this relation, we obtain
\bea
  \bm{a}_i 
  \lmk
 \bear{l}
  v_{l,p}^{\ q}(nT) \\
  \dot{v}_{l,p}^{\ q}(nT) \\
 \eear
 \rmk
 &=& 
  \bm{a}_i 
 \bm{M}(nT)
 \lmk
 \bear{ll}
  \frac{1}{\sqrt{2 \omega}} \bm{1}_{N \times N} \\
  \frac{-i\omega}{\sqrt{2 \omega}} \bm{1}_{N \times N} \\
 \eear
 \rmk, \nonumber\\
 &=&
 e^{n \mu_i T}
   \bm{a}_i 
 \lmk
 \bear{ll}
  \frac{1}{\sqrt{2 \omega}} \bm{1}_{N \times N} \\
  \frac{-i\omega}{\sqrt{2 \omega}} \bm{1}_{N \times N} \\
 \eear
 \rmk, \nonumber\\
 &=&
 e^{n \mu_i T} 
   \bm{a}_i 
  \lmk
 \bear{l}
  v_{l,p}^{\ q}(0) \\
  \dot{v}_{l,p}^{\ q}(0) \\
 \eear
 \rmk.
\eea
Therefore the linear combination of the mode functions with coefficients $\bm{a}_i$ grows exponentially
with the rate $\text{Re}[ \mu_i]$,
and this indicates that the $\chi$-particles  are produced exponentially.

%%%%%%%%%%%%%%%%%%%%%%%%%%%%%%%%%%%%%%%%%%%%%%%%%%%%%%%%%%
\section{\label{sec4}Results and physical meaning}
%%%%%%%%%%%%%%%%%%%%%%%%%%%%%%%%%%%%%%%%%%%%%%%%%%%%%%%%%%

We numerically calculate the maximum growth rate $\text{Re} [\mu]_{max}$
(hereafter we denote it as $\mu$ for simplicity).
We set $\mphi L / \pi \approx N \approx (30-60)$
for each $R$ and confirm that our results are independent of the size of volume $L$ within 1\%.
Hereafter, we ignore the small difference between $\omega_0$ and $m_\phi$, which is $O(K)$,
and simply set $\omega_0 = m_\phi$.

%%%%%%%%%%%%%%%%%%%%%%%%%%%%%%%%%%%%%%%%%%%%%%%%%%%%%%%%%%
\subsection{\label{sub1}Case of $m_\chi = 0$ and $l=0$}
%%%%%%%%%%%%%%%%%%%%%%%%%%%%%%%%%%%%%%%%%%%%%%%%%%%%%%%%%%

\begin{figure}[t]
\centering % \begin{center}/\end{center} takes some additional vertical space
  \includegraphics[width=110mm]{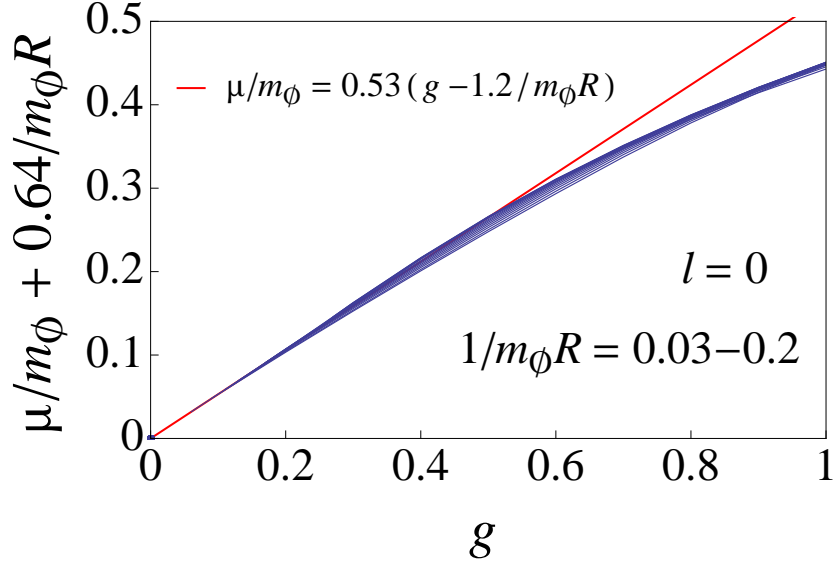}
  \caption{
  Maximum value of growth rate, $\mu$, shifted by the factor of $0.64 / R$
  as a function of coupling $g$ for $l=0$, $m_\chi = 0$, 
  and $1/(\mphi R) = 0.03 - 0.2$.
  Lines for $1/(\mphi R) = 0.03 - 0.2$ overlap and it is difficult 
  to distinguish among them.
  We fit the results by the red line for small $g$.
}
  \label{fig1}
\end{figure}

%%%%%%%%%%%%%%%%%%%%%%%%%%%%%ver.2%%%%%%%%%%%%%%%%%%%%%%%%%%%%%
%Figure~\ref{fig1} shows that
%the growth rate, $\mu$, shifted by $0.64 /(\mphi R)$ linearly depends on the coupling $g$ for small $g$.
Figure~\ref{fig1} shows that
the growth rate, $\mu$, shifted by $0.64 / R$ linearly depends on the coupling $g$ for small $g$.
%%%%%%%%%%%%%%%%%%%%%%%%%%%%%ver.2%%%%%%%%%%%%%%%%%%%%%%%%%%%%%
Before we make physical interpretation of this result,
let us review a resonance effect in preheating in the following.

%%%%%%%%%%%%%%%%%%%%%%%%%%%%%%%%%%%%%%%%%%%%%
%%%%%%%%%%%%%%%%%%%%%%%%%%%%%%%%%%%%%%%%%%%%%

If we neglect the spatial dependence of the I-ball,
what we calculate here is a growth rate due to parametric resonance 
in preheating~\cite{preheating}.
In this case, the equation of motion is given by
\beq
 \ddot{\chi} - \nabla^2 \chi + m_\chi^2 \chi 
 + g \omega_0^2 \cos (\omega_0 t) \chi = 0. 
\eeq
When we consider a fixed Fourier mode of $\chi$, 
this equation is reduced to the Mathieu equation as 
\beq
 \frac{\del^2}{\del z^2} \chi_k + \lmk A_k - 2q \cos 2z \rmk \chi_k = 0, 
\eeq
where
\bea
 A_k &=& 4 \frac{k^2 + m_\chi^2}{\omega_0^2}, \\
 q &=& 2 g, \\
 z &=& \frac{\omega_0 t}{2} + \frac{\pi}{2}. 
\eea
In the case of $q \lesssim 1$,
the resonance occurs in some narrow bands near $A_k \simeq 1^2, 2^2, 3^2, \dots$.
The most important instability band is the first one, $A_k \simeq 1$,
and the maximum growth rate is given by 
\beq
 \label{eq:growth_rate}
\mu_0 \simeq g \mphi / 2.
\eeq
The width of the first instability band is of the order of $q$.

The above resonance effect can be interpreted as the collective decay process by Bose stimulation.
The elementary decay rate of the field $\phi$ through the interaction of Eq.~(\ref{interaction}) is calculated as
\beq
 \label{eq:gamma_phi}
 \Gamma_\phi = \frac{g^2 m_\phi^2}{16 \pi \Phi_0^2} p_\chi,
\eeq
where $p_\chi = \sqrt{E_\chi^2 - m_{\chi, \text{eff}}^2(t)}$ and $E_\chi = \mphi/2$.
Since the mass of $\chi$ changes with time as $m_{\chi, \text{eff}} (t)^2 = m_\chi^2 + g \omega_0^2 \cos(\omega_0 t)$,
there is the uncertainty in the momentum of the field $\chi$ which is estimated as
\beq
 \Delta p_\chi = \frac{\Delta m_{\chi, \text{eff}}^2}{2p_\chi} \sim \frac{g \omega_0^2}{p_\chi}.
 \label{uncertainty}
\eeq
In order to take the effects of Bose enhancement into account,
we need to count the number of states where the $\chi$-particles can occupy. 
In our case, this can be estimated as
\bea
 \Delta N_{\rm ns} &\approx& \frac{\Delta^3 p_\chi \Delta^3 x}{(2 \pi)^3}, \nonumber\\
 &\approx& \frac{p_\chi^2 \Delta p_\chi V}{2 \pi^2}
 \approx \frac{g \omega_0^2 p_\chi V}{2 \pi^2},
\eea
where $V$ is an arbitrary scale of volume.
Then, we can estimate the production rate for each state 
from the decay of condensation $\phi$ as
\bea
  \label{eq:growth_rate2}
 \frac{1}{\Delta N_{\rm ns}} \lmk n_\phi V \rmk 2 \Gamma_\phi
 \approx \frac {2 \pi^2}{g \omega_0^2 p_\chi V} \lmk \half \omega_0 \Phi_0^2 V \rmk 
 \frac{2 g^2 m_\phi^2}{16 \pi \Phi_0^2} p_\chi
 \approx \frac{\pi}{8} g m_\phi,
\eea
where we use $\omega_0 \simeq m_\phi$,
and $n_\phi$ ($= \omega_0 \Phi_0^2/2$) is the number density of the field $\phi$.
We insert the factor of $2$ because two particles of $\chi$ are produced through each decay of $\phi$.
This result is independent from the arbitrary scale of volume $V$, as expected.
Since the actual decay rate is proportional to the occupation number
in the final state
due to Bose stimulation,
the number density of particles $\chi$ grows exponentially with the growth rate of $\mu_0 \approx \pi g m_\phi /8$.
This result is consistent with Eq.~(\ref{eq:growth_rate}) which is derived 
on the basis of the Mathieu equation.

In the case of the I-ball with finite radius $R$ ($= 1/K^{1/2} m_\phi$),
the particles in the final state escape from the I-ball.
Therefore
the effects of Bose enhancement are relevant
only if the particle production rate is larger than a certain escape rate 
from the I-ball~\cite{previous work}.
The escape rate can be estimated as $v_\chi / R$ where $v_\chi$ is
the velocity of the particle $\chi$,
and we can say that the I-ball decay rate is affected by Bose enhancement
when $\mu_0 > \mu_* \sim v_\chi/ R$.
In fact, Fig.~\ref{fig1} shows that
the growth rate is approximately given by
\beq
 \mu \simeq 0.53 \mphi \lmk g - \frac{1.2}{\mphi R} \rmk,
\eeq
for small $g$ and $m_\chi =0$.
This indicates that $\mu_* \simeq 0.64 /R$ for $m_\chi = 0$.
Also, the magnitude of the proportionality constant ($0.53 \mphi$) 
is consistent with the analytic estimation derived above 
[see Eq.~(\ref{eq:growth_rate}) or (\ref{eq:growth_rate2})].
When we define $g^*$ as a critical value of the coupling constant
above which a non-zero growth rate $\mu$ is obtained,
it is given by $g^* \simeq 1.2/(\mphi R)$ for $m_\chi =0$.

%%%%%%%%%%%%%%%%%%%%%%%%%%%%%%%%%%%%%%%%%%%%%%%%%%%%%%%%%%
\subsection{\label{sub2}Case of $m_\chi > 0$ and $l=0$}
%%%%%%%%%%%%%%%%%%%%%%%%%%%%%%%%%%%%%%%%%%%%%%%%%%%%%%%%%%

%%%%%%%%%%%%%%%%%%%%%%%%%%%%%%%%%%%%%%%%%%%%%
\begin{figure*}
\centering 
\includegraphics[width=.45\textwidth]{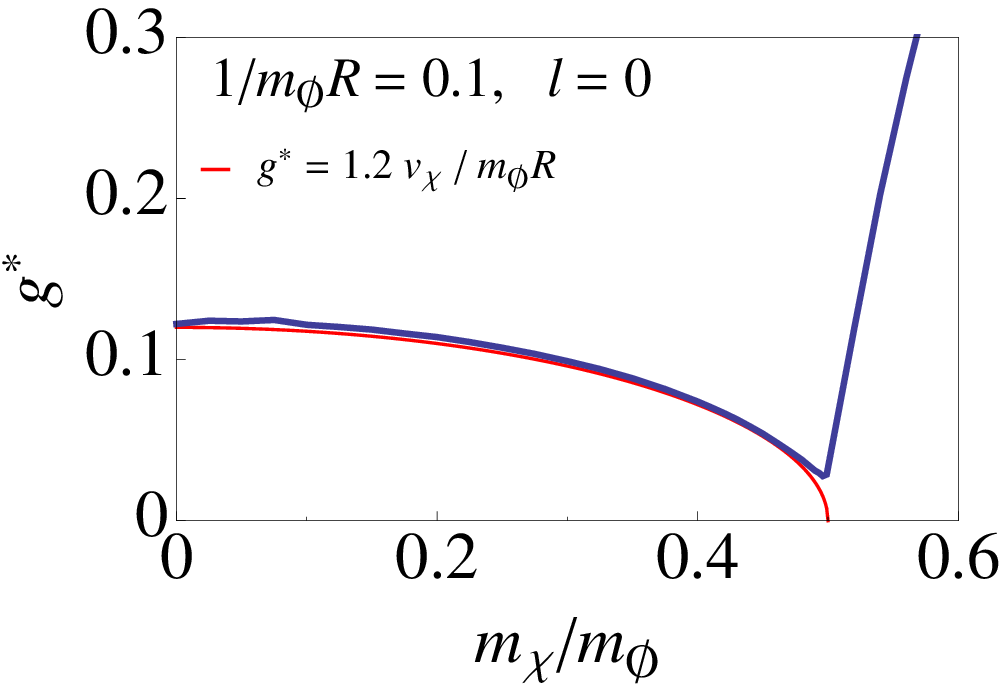} 
\hfill
\includegraphics[width=.45\textwidth]{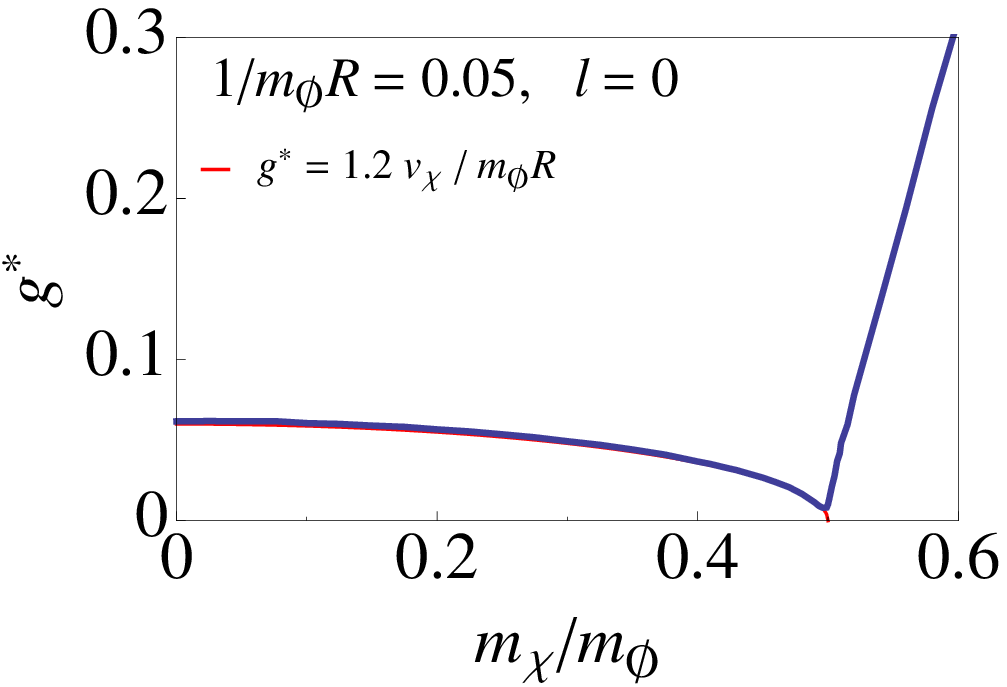} 
  \caption{
  Critical value of the coupling $g^*$
  as a function of mass $m_\chi$ for $l=0$.
  The inverse of radius is $1/(\mphi R) = 0.1$ (left) and $0.05$ (right).
  We fit the results by red curves in light of $g^* \propto v_\chi$ for $m_\chi \lesssim \mphi/2$.
  Critical values of the coupling $g^*$ reach finite lower bounds for $m_\chi \approx \mphi/2$
  because of the uncertainty in the momentum estimated in Eq.~(\ref{uncertainty}).
}
  \label{fig2}
\end{figure*}
%%%%%%%%%%%%%%%%%%%%%%%%%%%%%%%%%%%%%%%%%%%%%

Next, we study the effect of a finite mass of $\chi$.
Figure~\ref{fig2} shows the mass dependence of $g^*$ as
\beq
 g^* \propto v_\chi = \frac{p_\chi}{E_\chi} 
 \quad \text{ for }\  m_\chi \lesssim \frac{\mphi}{2},
\eeq
where
$p_\chi = \sqrt{E_\chi^2 - m_\chi^2}$
and 
$E_\chi = \mphi /2$.
This is consistent with the explanation given in the previous subsection 
that the escape rate is proportional to the velocity of $\chi$.
However, there is a lower bound for the velocity
since there is the uncertainty in the momentum of the field $\chi$,
which is estimated in Eq.~(\ref{uncertainty}).
Using Eq.~(\ref{uncertainty}) with $p_\chi \sim \Delta p_\chi$ for $m_\chi \approx \mphi/2$,
we can estimate the lower bound of the effective velocity as $\text{Min}[v_\chi] \sim 2g^{1/2}$.
Therefore the critical value of the coupling $g^*$ reaches a non-zero lower bound $g^*_{min}$
when $m_\chi \approx \mphi/2$,
where
\bea
  g^*_{\text{min}} &\sim & \frac{\text{Min}[v_\chi] }{\mphi R} \sim \frac{2 (g^*_{\text{min}})^{1/2}}{\mphi R},\nonumber \\
 \Leftrightarrow ~~~~~ g^*_{\text{min}} &\sim & \frac{4}{(\mphi R)^2}.
  \label{eq:g_min}
\eea
We thus obtain the analytic estimation for $g^*_{\text{min}}$, which can be 
compared with the result shown in Fig.~\ref{fig2}.
In fact, the figure indicates that numerically it is given by
\beq
 g^*_{\text{min}} \simeq \frac{2}{(\mphi R)^2},
 \label{g lower bound}
\eeq
which is consistent with Eq~(\ref{eq:g_min}).

%%%%%%%%%%%%%%%%%%%%%%%%%%%%%%%%%%%%%%%%%%%%%%%%%%%%%%%%%%
\subsection{\label{sub3}Case of $m_\chi = 0$ and $l > 0$}
%%%%%%%%%%%%%%%%%%%%%%%%%%%%%%%%%%%%%%%%%%%%%%%%%%%%%%%%%%

%%%%%%%%%%%%%%%%%%%%%%%%%%%%%%%%%%%%%%%%%%%%%
\begin{figure}
\centering 
  \includegraphics[width=110mm]{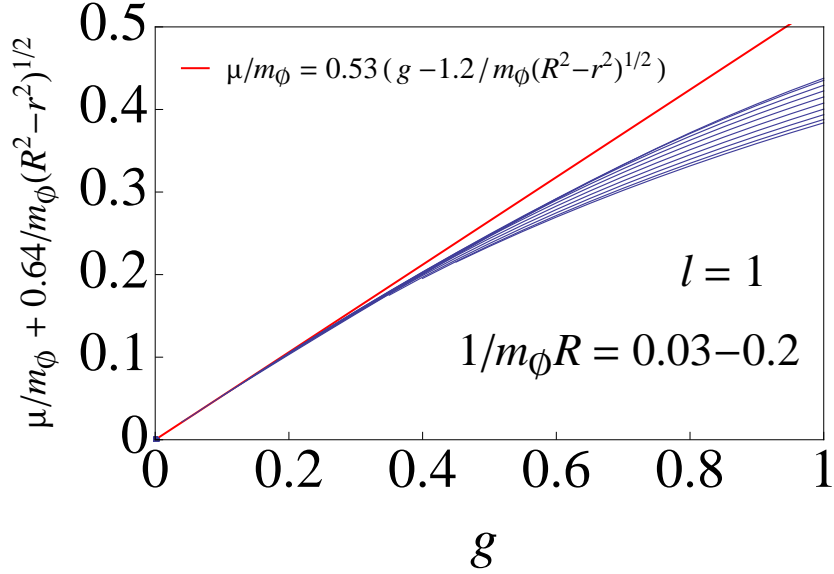}
  \caption{
  Maximum value of growth rate, $\mu$, shifted by the factor of $0.64 / (R^2-r^2)^{1/2}$
  as a function of coupling $g$ for $l=1$, $m_\chi = 0$, and $1/(\mphi R) = 0.03 - 0.2$,
  where $r = [l(l+1)]^{1/2}/p_\chi$.
  We fit the results by the red line for small $g$.
}
  \label{fig3}
\end{figure}
%%%%%%%%%%%%%%%%%%%%%%%%%%%%%%%%%%%%%%%%%%%%%

\begin{figure}
\centering 
  \includegraphics[width=85mm]{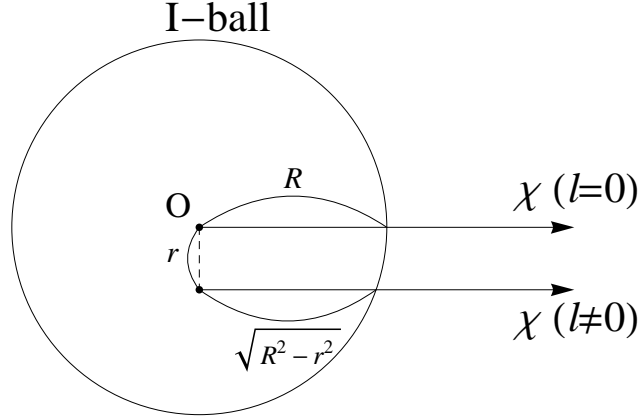}
  \caption{
  Classical description for the motion of $\chi$
  produced from I-ball decay.
  For $l \ne 0$, the daughter particle $\chi$
  is produced at a finite distance from the origin, $r$.
%%%%%%%%%%%%%%%%%%%%%%%%%%%%%ver.2%%%%%%%%%%%%%%%%%%%%%%%%%%%%%
%  The distance from that point to the I-ball surface is given by
%  $R$ and $\sqrt{R^2 - r^2}$ for $l=0$ and $l \ne 0$, respectively.
  The distance from that point to the I-ball surface is given by
  $\sqrt{R^2 - r^2}$.
%%%%%%%%%%%%%%%%%%%%%%%%%%%%%ver.2%%%%%%%%%%%%%%%%%%%%%%%%%%%%%
}
  \label{fig4}
\end{figure}

Let us consider the growth rate when the daughter $\chi$-particle has 
angular momentum $\ell$.
Figure~\ref{fig3} shows that growth rates depend complicatedly on the size 
of I-ball $R$ for the case of $l=1$.
These behaviors are understood by taking the finite-size effect into account.
In classical mechanics, the angular momentum of a particle is given by the product of its momentum, $p_\chi$,
and its position from the origin, $r$.
When a particle is produced at some position away from the origin,
the distance from that point to the I-ball surface is given by $\sqrt{R^2-r^2}$ 
(see Fig.~\ref{fig4}).
Therefore its escape rate is now given by
\beq
 \frac{v_\chi}{\mphi \sqrt{R^2-r^2}}.
\eeq
In the case considered here, $r = [l(l+1)]^{1/2}/p_\chi$
since the total angular momentum is given by $[l(l+1)]^{1/2}$.
Taking these into account,
we fit the results in Fig.~\ref{fig3} as
\beq
 g^* \simeq \frac{1.2 v_\chi}{\mphi R \sqrt{1- l(l+1) / (p_\chi R)^2}}.
 \label{fit formula}
\eeq
%%

%%%%%%%%%%%%%%%%%%%%%%%%%%%%%%%%%%%%%%%%%%%%%
%%%%%%%%%%%%%%%%%%%%%%%%%%%%%%%%%%%%%%%%%%%%%

%%%%%%%%%%%%%%%%%%%%%%%%%%%%%%%%%%%%%%%%%%%%%%%%%%%%%%%%%%
\subsection{\label{sub4}Case of $m_\chi > 0$ and $l > 0$}
%%%%%%%%%%%%%%%%%%%%%%%%%%%%%%%%%%%%%%%%%%%%%%%%%%%%%%%%%%

%%%%%%%%%%%%%%%%%%%%%%%%%%%%%%%%%%%%%%%%%%%%%
\begin{figure*}[!h]
\centering 
\includegraphics[width=.45\textwidth]{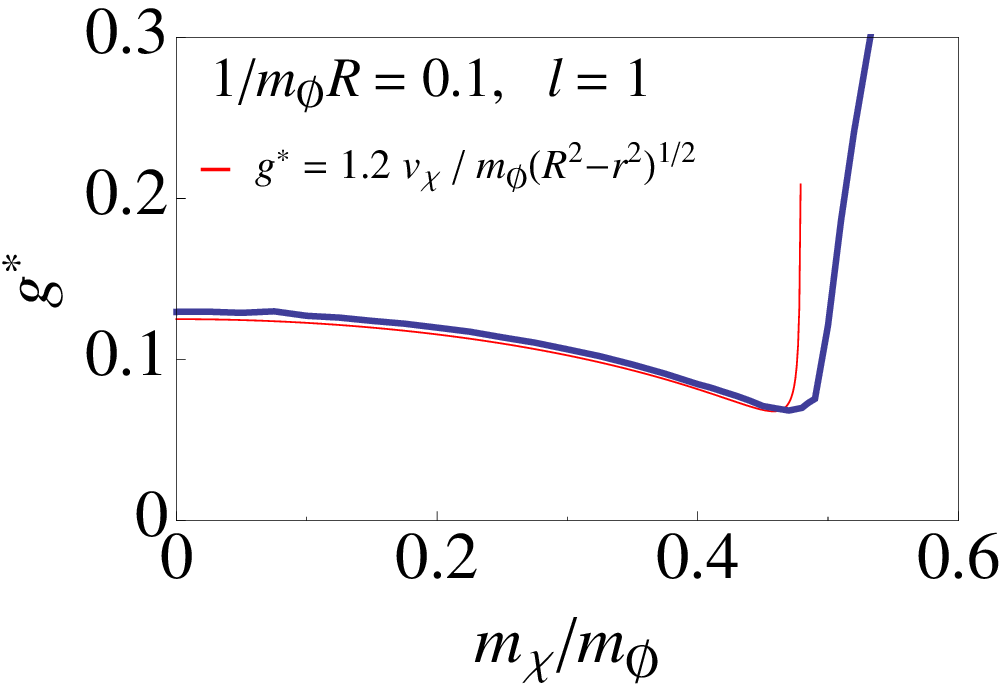} 
\hfill
\includegraphics[width=.45\textwidth]{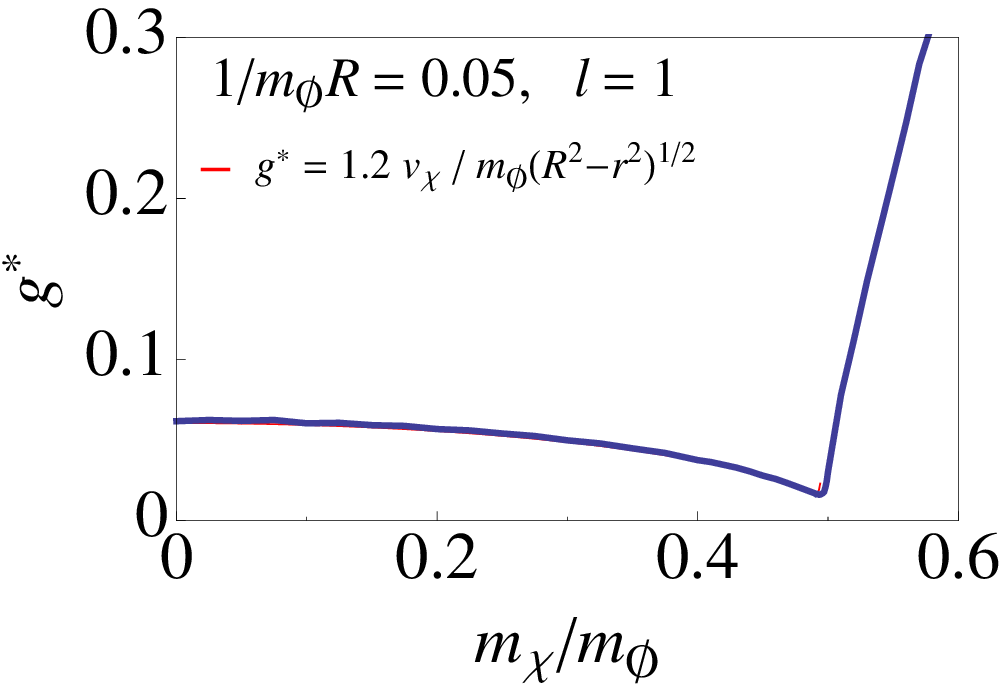} \\
\vspace{1cm}
\includegraphics[width=.45\textwidth]{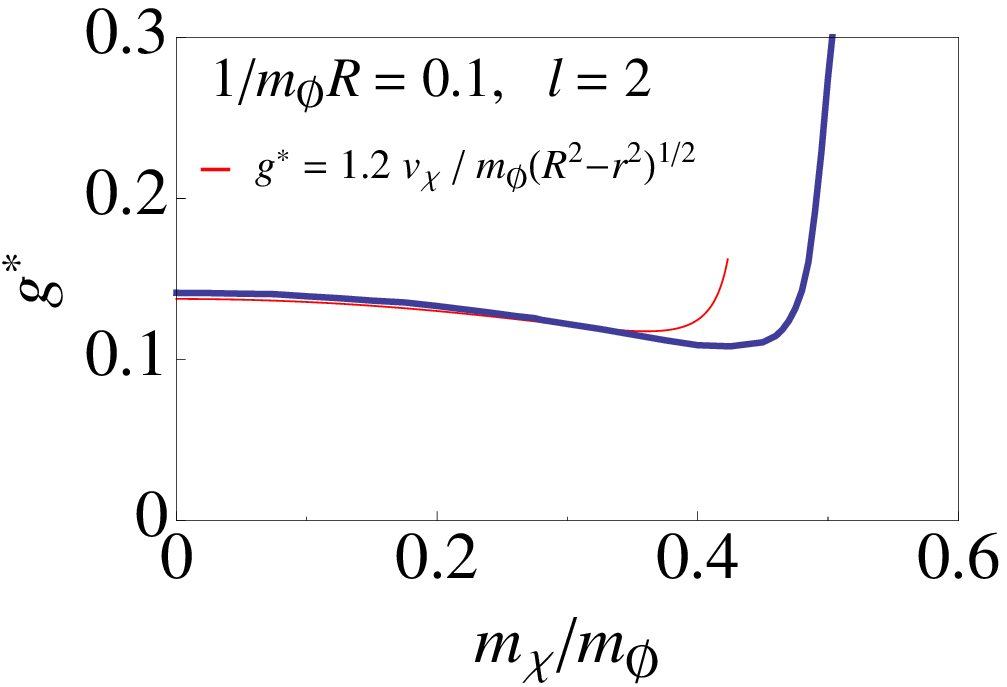} 
\hfill
\includegraphics[width=.45\textwidth]{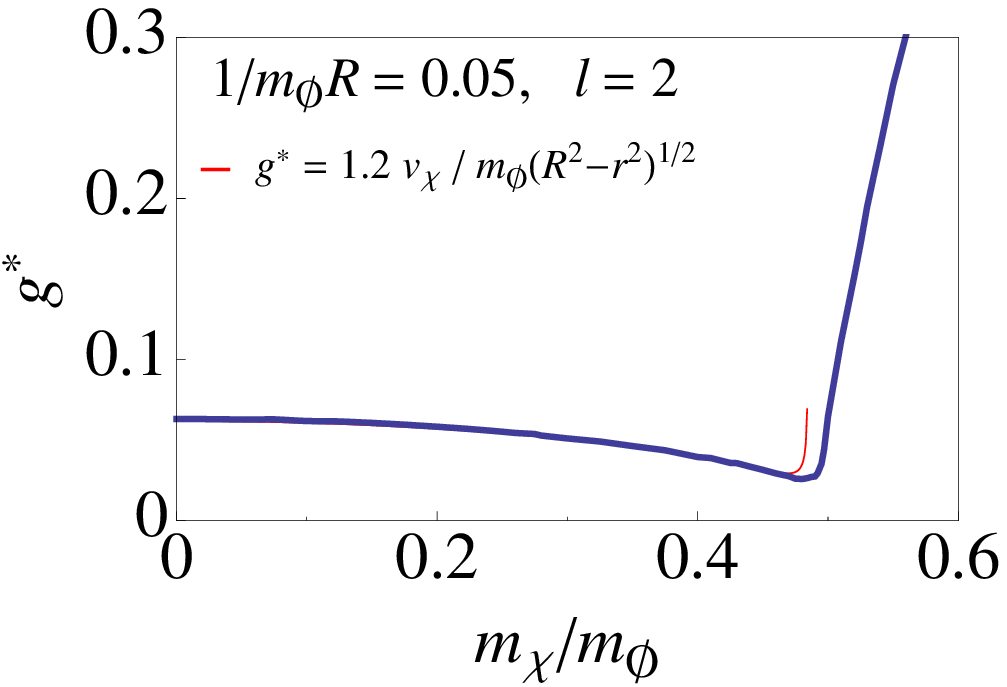} \\
\vspace{1cm}
\includegraphics[width=.45\textwidth]{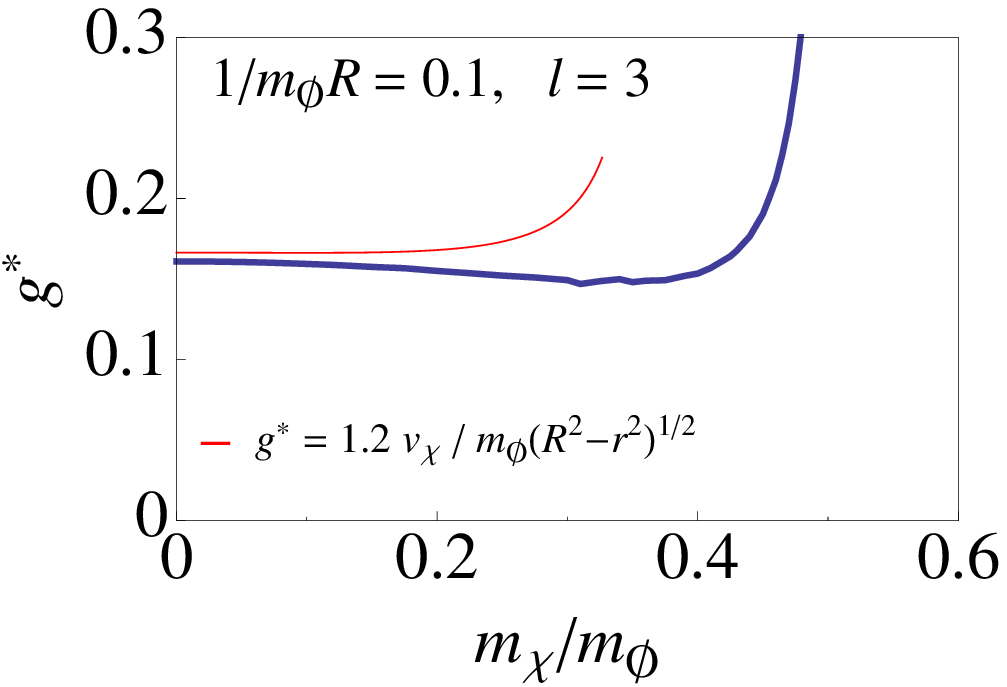} 
\hfill
\includegraphics[width=.45\textwidth]{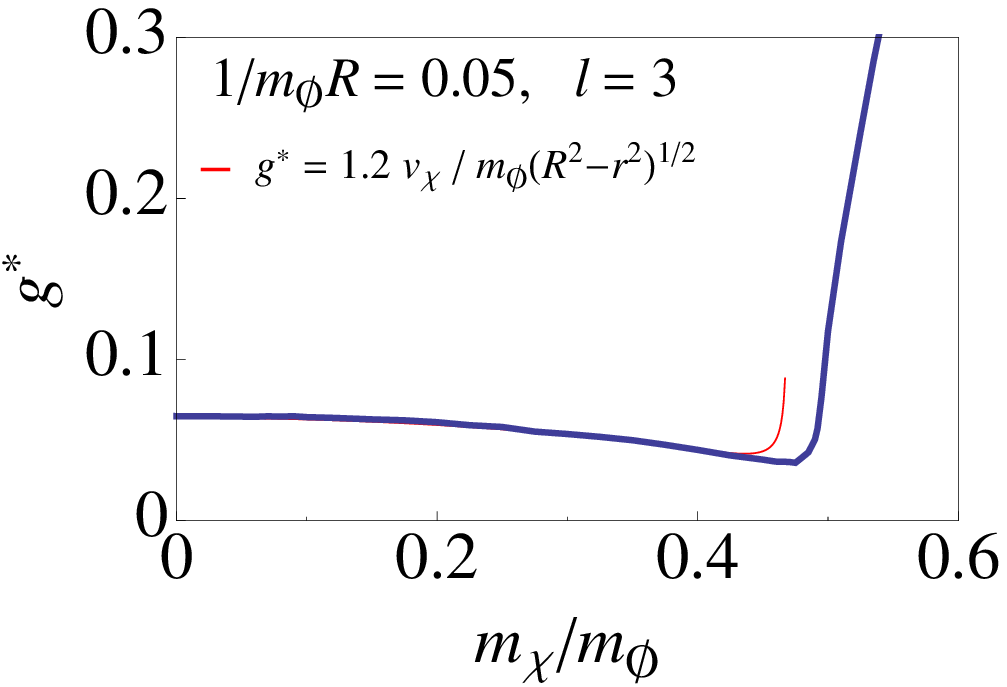} 
  \caption{
  Critical value of the coupling $g^*$
  as a function of mass $m_\chi$ for $l=1$ (top), $2$ (middle), and $3$ (bottom).
  The inverse of radius is $1/(\mphi R) = 0.1$ (left) and $0.05$ (right).
  Red curves are functions derived analytically as 
  $g^* = 1.2 v_\chi/ \mphi (R^2 - r^2)^{1/2}$.
  There are gaps between the numerical and analytical results
  due to the uncertainty in the momentum, which is proportional to the coupling $g$ (see Eq.~(\ref{uncertainty})).
}
  \label{fig5}
\end{figure*}
%%%%%%%%%%%%%%%%%%%%%%%%%%%%%%%%%%%%%%%%%%%%%

We also calculate the mass dependence of $g^*$ for each $l$ ($=1$, $2$, and $3$) as shown in Fig.~\ref{fig5}.
The results are consistent with Eq.~(\ref{fit formula})
when we take into account the uncertainty in the momentum estimated in Eq.~(\ref{uncertainty}).
Since $v_\chi = p_\chi / E_\chi$ and $E_\chi = m_\phi/2$,
we can find the lower bound for $g^*$ from Eq.~(\ref{fit formula})
by carrying out the differentiation and set it equal to zero.
If we neglect the uncertainty in the momentum,
we obtain
\beq
 \text{Min} \lkk g^* \rkk = 1.2 \times \frac{3 \sqrt{3l(l+1)}}{(\mphi R)^2},
 \label{g min for l}
\eeq
at the momentum of 
\beq
 p_\chi = \frac{ 3\sqrt{l(l+1)}}{2R}.
 \label{p chi1}
\eeq
If this momentum is sufficiently larger than the uncertainty in the momentum,
the derivation of Eqs.~(\ref{g min for l}) and (\ref{p chi1}) is consistent.
On the other hand,
if the momentum of Eq.~(\ref{p chi1}) is less than the uncertainty in the momentum,
the lower bound for $g^*$ is again determined by Eq.~(\ref{g lower bound}).
Equation~(\ref{g min for l}) indicates
that the lower bound of $g^*$ increases with increasing the angular momentum $l$,
as we can see in Fig.~\ref{fig5}.

%%%%%%%%%%%%%%%%%%%%%%%%%%%%%%%%%%%%%%%%%%%%%%%%%%%%%%%%%%
%%%%%%%%%%%%%%%%%%%%%%%%%%%%%ver.2%%%%%%%%%%%%%%%%%%%%%%%%%%%%%
%\subsection{\label{sub5}Summary and discussions}
\section{\label{discussions}Summary and discussions}
%%%%%%%%%%%%%%%%%%%%%%%%%%%%%ver.2%%%%%%%%%%%%%%%%%%%%%%%%%%%%%
%%%%%%%%%%%%%%%%%%%%%%%%%%%%%%%%%%%%%%%%%%%%%%%%%%%%%%%%%%

%%%%%%%%%%%%%%%%%%%%%%%%%%%%%ver.2%%%%%%%%%%%%%%%%%%%%%%%%%%%%%
%Our results are summarized as follows.
Our results calculated in section~\ref{sec4} are summarized as follows.
%%%%%%%%%%%%%%%%%%%%%%%%%%%%%ver.2%%%%%%%%%%%%%%%%%%%%%%%%%%%%%
If we neglect the uncertainty in the momentum,
the growth rate is approximately written as
\beq
 \mu \simeq 0.53 \mphi \lmk g - g^* \rmk \qquad \text{ for } g \ll 1,
\label{growth rate}
\eeq
%%
%%%%%%%%%%%%%%%%%%%%%%%%%%%%%ver.2%%%%%%%%%%%%%%%%%%%%%%%%%%%%%
%where $g^*$ is given by Eq.~(\ref{fit formula}).
%When $l=0$ and $m_\chi \approx \mphi/2$,
%the uncertainty in the momentum given by Eq.~(\ref{uncertainty}) is important 
%and $g^*$ approaches the value of Eq.~(\ref{g lower bound}).
where $g^*$ is given by
\beq
 g^* \simeq \frac{1.2 v_\chi}{\mphi R \sqrt{1- l(l+1) / (p_\chi R)^2}}.
 \label{fit formula2}
\eeq
When $l=0$ and $m_\chi \approx \mphi/2$,
the uncertainty in the momentum given by Eq.~(\ref{uncertainty}) is important
and $g^*$ approaches the value of
\beq
 g^*_{\text{min}} \simeq \frac{2}{(\mphi R)^2}.
 \label{g lower bound2}
\eeq
%%
%%%%%%%%%%%%%%%%%%%%%%%%%%%%%ver.2%%%%%%%%%%%%%%%%%%%%%%%%%%%%%
As apparent from Eq.~(\ref{fit formula2}), the critical value is larger for 
non-zero angular momentum.
In other words, the rate of exponential decay with non-zero angular momentum
is smaller than the one with zero angular momentum.

Since the decay rate of I-balls grows exponentially as $\sim \exp(\mu t)$,
I-balls decay instantaneously at the time of $t \sim \mu^{-1} \sim H^{-1}$, where $H$ is the Hubble parameter.
From Eq.~(\ref{growth rate}), the temperature at the I-ball decay, $T_\text{d}$, can be calculated as
\begin{eqnarray}
  T_{\text{d}} 
  & \simeq & \left( \frac{90}{4 \pi^2 g_s} \right)^{1/4} \sqrt{\mu M_\text{P}}, 
  \nonumber\\
  &\simeq& 10^{10}\text{ GeV} \times (g-g^*)^{1/2} \left( \frac{\mphi}{\text{TeV}} \right)^{1/2}, \label{decay temp}
\end{eqnarray} 
where $g_s$ is the effective relativistic degrees of freedom at the decay time.
%%%%%%%%%%%%%%%%%%%%%%%%%%%%%ver.2%%%%%%%%%%%%%%%%%%%%%%%%%%%%%
This reheating temperature is larger than the one derived from
the perturbative decay rate (\ref{eq:gamma_phi})
by the factor of
\begin{eqnarray}
  \lmk \frac{\mu}{\Gamma_\phi} \rmk^{1/2} \simeq \lmk 0.53 \frac{g-g^*}{g} \frac{16 \pi \Phi_0^2}{ g \mphi p_\chi} \rmk^{1/2},
\end{eqnarray} 
where we use Eq.~(\ref{eq:gamma_phi}).
This factor is larger for larger I-balls (see Eq.~(\ref{phi0})) and for smaller coupling constant.
Another important property is that
I-balls can decay only into the particle which interacts with the I-balls with a coupling constant larger than $g^*$.
This property may allow us to obtain a high reheating temperature without producing unwanted relics
and may lead to new cosmological scenarios of non-thermal dark matter production and non-thermal leptogenesis.
%%%%%%%%%%%%%%%%%%%%%%%%%%%%%ver.2%%%%%%%%%%%%%%%%%%%%%%%%%%%%%

Finally, we comment on the difference between preheating and I-ball decay.
Since the growth rate is proportional to the amplitude of the oscillating field $\Phi_0$,
it decreases due to the Hubble expansion as $\mu \propto \Phi_0 \propto a^{-3/2}$
in the case of preheating, where $a$ is the scale factor.
On the other hand, since the dynamics of the amplitude of the I-ball $\phi_0$ decouples from the Hubble expansion,
the growth rate $\mu$ remains constant in time in the case of I-ball decay.
Therefore the I-ball eventually decays exponentially when $\mu > 0$,
i.e., when the coupling of interaction is larger than the critical value of the coupling, $g^*$.

%%%%%%%%%%%%%%%%%%%%%%%%%%%%%%%%%%%%%%%%%%%%%%%%%%%%%%%%%%
\section{\label{conclusion}Conclusions}
%%%%%%%%%%%%%%%%%%%%%%%%%%%%%%%%%%%%%%%%%%%%%%%%%%%%%%%%%%

We have focused on I-balls in the scalar field theory
with the monomial potential of $V(\phi) \propto \phi^{2(1-K)}$ and $0< K \ll 1$ in $3+1$ dimensions,
which is motivated by chaotic inflationary models and supersymmetric theories.
The stability of I-ball is guaranteed by the conservation of an adiabatic invariant,
which also determines the configuration of the I-ball as Gaussian in that theory.
We have calculated the decay rate of the Gaussian-type I-ball
through a interaction with another scalar field, taking into account the effects of Bose enhancement.
%%%%%%%%%%%%%%%%%%%%%%%%%%%%%ver.2%%%%%%%%%%%%%%%%%%%%%%%%%%%%%
%and have revealed the conditions that the I-ball decays exponentially.
In Ref.~\cite{previous work}, a non-perturbative method to compute the decay rate has been derived in general dimensions.
We have applied the method assuming spherical symmetry
and have calculated the partial decay rates into partial waves, labelled by the angular momentum
of daughter particles.
We have also revealed the conditions that the I-ball decays exponentially.
%%%%%%%%%%%%%%%%%%%%%%%%%%%%%ver.2%%%%%%%%%%%%%%%%%%%%%%%%%%%%%

While the effects of Bose enhancement is proportional to the number 
density of the daughter particles inside the I-ball,
the daughter particles escape from the I-ball.
Therefore
the effects of Bose enhancement are relevant
only if a production rate is larger than a certain escape rate from the I-ball~\cite{previous work}.
In other words, there is a critical value of the coupling constant
%%%%%%%%%%%%%%%%%%%%%%%%%%%%%ver.2%%%%%%%%%%%%%%%%%%%%%%%%%%%%%
%above which the I-ball decay exponentially.
above which the I-ball decays exponentially due to Bose stimulation
and below which it decays linearly through elementary decay processes.
%%%%%%%%%%%%%%%%%%%%%%%%%%%%%ver.2%%%%%%%%%%%%%%%%%%%%%%%%%%%%%
The critical value is basically proportional to the product of the velocity 
of daughter particles and the inverse of the size of the I-ball.
However,
we have to take account of the lower bound of velocity,
which comes from an uncertainty in the momentum of daughter particles
and is again proportional to the inverse of the size of the I-ball.
In the case of non-zero angular momentum for daughter particles,
the critical value is larger than the one in the case of zero angular momentum.
Therefore, the rate of exponential decay with non-zero angular momentum
is smaller than the one with zero angular momentum.

%%%%%%%%%%%%%%%%%%%%%%%%%%%%%ver.2%%%%%%%%%%%%%%%%%%%%%%%%%%%%%
In chaotic inflation models with the above potential,
I-balls are in fact formed under some conditions~\cite{oscillon2011}.
In this scenario,
inflaton begins to oscillate soon after inflation ends, and then instabilities of the inflaton oscillation grow to form I-balls.
Since I-balls still dominate the energy density of the Universe,
the decay rate of I-ball determines the reheating temperature of the Universe.
If the decay rate of I-ball is enhanced by Bose stimulation,
the reheating temperature is much larger than the one derived from
the perturbative decay rate.
Another important consequence is that
I-balls can decay only into the particle which interacts with the I-balls with coupling constants larger than the critical value.
These properties may lead to some implications for the physics related to the reheating process of the Universe.
%%%%%%%%%%%%%%%%%%%%%%%%%%%%%ver.2%%%%%%%%%%%%%%%%%%%%%%%%%%%%%

%%%%%%%%%%%%%%%%%%%%%%%%%%%%%%%%%%%%
\acknowledgments
%%%%%%%%%%%%%%%%%%%%%%%%%%%%%%%%%%%%
This work was supported by Grant-in-Aid for Scientific research
from the Ministry of Education, Science, Sports and Culture
(MEXT), Japan, No. 25400248 (M.K.), No. 21111006 (M.K.),
by World Premier International Research Center Initiative
(WPI Initiative), MEXT, Japan (M.K.),
JSPS Research Fellowship for Young Scientists (M.Y.),
and the Program for Leading Graduate Schools, MEXT, Japan (M.Y.).

%%%%%%%%%%%%%%%%%%%%%%%%%%%%%%%%%%%%%%%%%%%%%%%%%%%%%%%%%%%%%%%%%%%%%%%%%%%%

%%%%%%%%%%%%%%%%%%%%%%%%%%%%%%%%%%%%

%%%%%%%%%%%%%%%%%%%%%%%%%%%%%%%%%%%%

\end{document}